\documentclass[twocolumn,prl,amsmath,letterpaper,floatfix,showpacs]{revtex4}
\usepackage{graphicx}
\usepackage{amsmath}
\usepackage{amssymb}
\usepackage{color}
\usepackage{gensymb}

\newcommand{\otwo}[0]{$\mathrm{O_2}$}
\newcommand{\mub}[0]{\mu_\mathrm{B}}
\newcommand{\Ml}[0]{M_\parallel}
\newcommand{\Mt}[0]{M_\perp}
\newcommand{\Mul}[0]{\mu_\parallel}
\newcommand{\Mut}[0]{\mu_\perp}
\newcommand{\Ea}[0]{\mathcal{E}}

\begin{document}

    \title{Ultrafast magnetization of a dense molecular gas with an optical centrifuge}

    \author{A.~A.~Milner}
    \author{A.~Korobenko}
    \author{V.~Milner}
    \date{\today}

    \affiliation{Department of  Physics \& Astronomy, The University of British Columbia, Vancouver, Canada}

    \begin{abstract}
Strong laser-induced magnetization of oxygen gas at room temperature and atmospheric pressure is achieved experimentally on the sub-nanosecond time scale. The method is based on controlling the electronic spin of paramagnetic molecules by means of manipulating their rotation with an optical centrifuge. Spin-rotational coupling results in high degree of spin polarization on the order of one Bohr magneton per centrifuged molecule. Owing to the non-resonant interaction with the laser pulses, the demonstrated technique is applicable to a broad class of paramagnetic rotors. Executed in a high-density gas, it may offer an efficient way of generating macroscopic magnetic fields remotely (as shown in this work), producing large amount of polarized electrons and converting electronic to nuclear spin polarization.

    \end{abstract}

    \maketitle

\section{Introduction}
Numerous studies in fundamental and applied sciences utilize gases with spin-polarized outer-shell electrons. Most notably, such gases are used for converting their electronic polarization to the polarization of nuclear spins \cite{Walker1997}, one of the key elements in nuclear magnetic resonance imaging \cite{Goodson2002}. Equally important, they provide a source of spin-polarized electrons \cite{Keliher1975, Gray1983} for particle-physics experiments \cite{Garcia2008}, as well as for probing molecular dynamics during chemical reactions and analyzing the electronic properties of materials \cite{Gay2009}.

At room temperature, lining up electronic spins by applying even the strongest laboratory-scale magnetic field is rather inefficient. In a widely used alternative approach, spin alignment is executed by means of optical pumping, in which an atom undergoes repetitive cycles of the resonant absorption of circularly polarized light, followed by spontaneous emission \cite{Happer1987}. Spontaneous life time dictates the time scale on the order of 100~ns for the high degree of spin polarization to be achieved with bound electrons. The process can be executed faster if the polarized electron is ejected from the parent atom by means of either frequency-resolved \cite{Lambropoulos1973, Nakajima2002} or time-resolved \cite{Bouchene2001, Nakajima2004} multi-photon ionization.

Since all of the above methods rely on the resonant atom-laser interaction, they are limited by the availability of strong light sources with the required wavelength. Alkali vapors -- the most common gases for which such sources are readily available -- have to be kept at low densities on the order of $10^{14}$~cm$^{-3}$ to prevent clusterization, strong absorption of pumping light and collisional depolarization \cite{Lancor2010, Walker2011}, thus providing relatively low amount of spin polarized electrons. High densities of rare gases, on the other hand, cannot be fully exploited because of the lack of high-intensity tunable vacuum ultra-violet sources, needed for efficient multi-photon ionization.

In this report, we demonstrate an alternative method of electronic spin polarization. The technique is based on the forced unidirectional rotation of molecules in an optical centrifuge \cite{Karczmarek1999, Villeneuve2000}. Unlike optical pumping, it operates  far off electronic resonances and is applicable to a broad class of Hund's case (b) molecules. Owing to its non-resonant nature, the process can be carried out in a dense gas -- here, in oxygen at atmospheric pressure. Even though only a few percent of \otwo{} molecules are centrifuged at room temperature, the observed density of polarized electrons exceeds $6\times10^{17}$~cm$^{-3}$. Such high degree of gas magnetization, three orders of magnitude above the magnetization typically obtained with optical pumping, corresponds to the centrifuge-induced magnetic field on the scale of tens of milligauss, detectable with a simple pickup coil. Finally, we show that the electronic polarization occurs on a sub-nanosecond time scale and can be enhanced, as well as accelerated, by placing the gas in a permanent external magnetic field of moderate strength.

\begin{figure}[ht]
    \begin{center}
    \includegraphics[width=1\columnwidth]{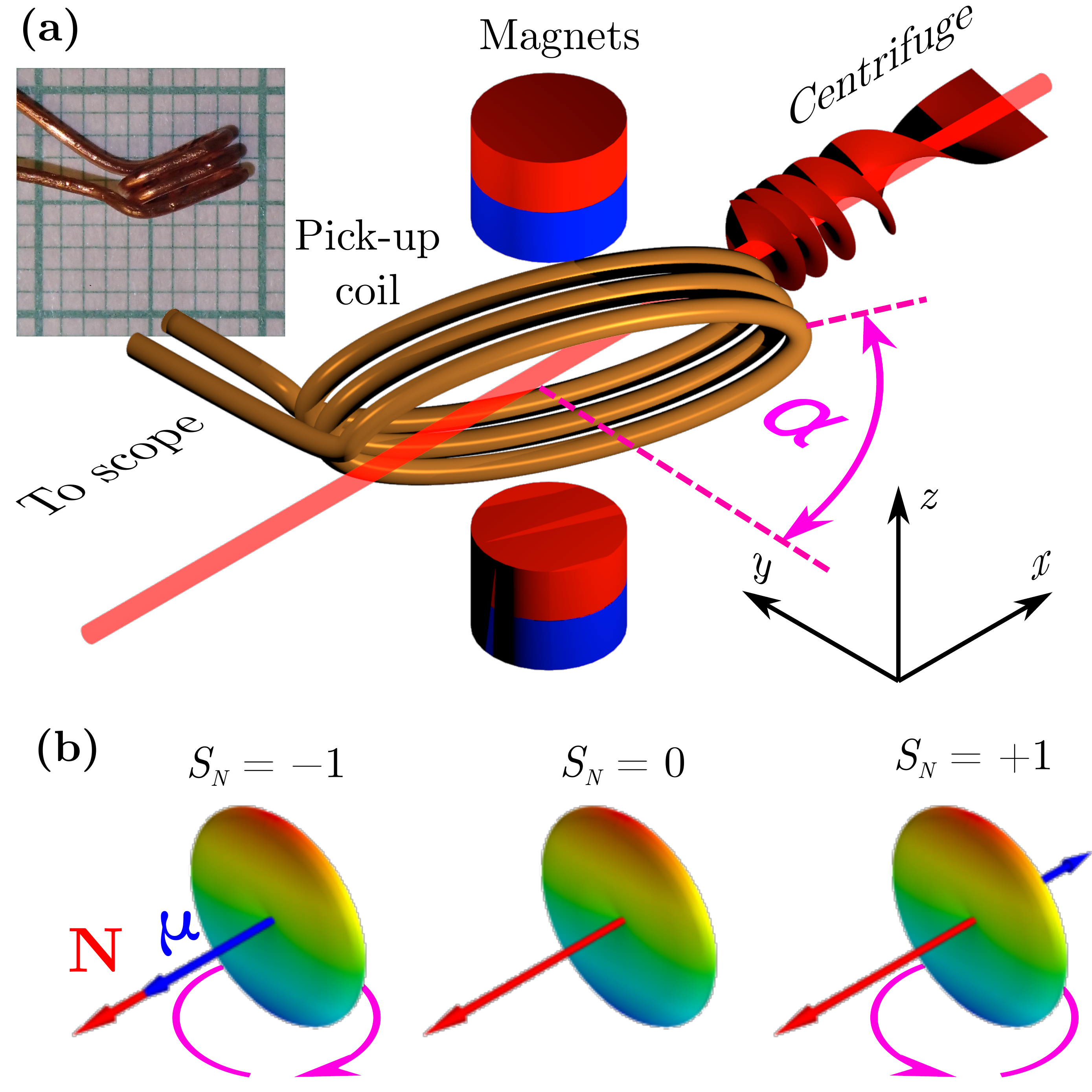}
    \caption{\label{fig_setup} (a) Experimental configuration. Centrifuge pulses, illustrated with a corkscrew-shaped red surface, pass through a pick-up coil between a pair of permanent magnets. The coil, pictured in the inset, was tilted by an angle $\alpha =59^{\circ}$,  had an elliptical cross-section of $0.93\times3.8$~mm$^2$, and was used for measuring the transverse magnetization (see text for details).  Another coil (not shown) with a circular cross-section (diameter of 1.2~mm) and its axis collinear with the centrifuge beam (thick red line) was used for detecting the longitudinal magnetization. (b) Molecular gyroscopes, represented by the disk-shaped distributions of molecular axes in the plane of rotation, with three possible projections of the electronic spin $\mathbf{\mu}$ (short blue arrows) on the molecular angular momentum $\mathbf{N}$ (long red arrows). Pink arrows under the left and right gyroscope show the direction of Larmor precession in the external magnetic field (see text).}
    \end{center}
\end{figure}
\section{Experimental setup}
Our optical setup has been described in detail elsewhere \cite{Milner2014b}. Briefly, the output beam of a broadband Ti:Sapphire laser (30~fs, 10~mJ at 800~nm) passes through a centrifuge pulse shaper, where the pulses are split into two spectral components, whose frequencies are linearly chirped in time. The applied chirps are equal in magnitude ($\approx 0.17~$THz/ps), but opposite in sign. The two spectral components are then separately amplified to $\approx 15$~mJ each in a home-built multi-pass amplifier, circularly polarized in opposite directions and combined in space and time to produce the field of an optical centrifuge, schematically shown in Fig.\ref{fig_setup}.

Centrifuge pulses were focused with a $f=1000~$mm lens inside a hermetic chamber, filled with oxygen at room temperature and variable pressure. To determine the degree of rotational excitation, we used state-resolved coherent Raman spectroscopy. The latter is executed by sending a weak narrowband probe pulse through the gas of centrifuged \otwo{} molecules and measuring the rotational Raman shift \cite{Korech2013, Korobenko2014a}. Our centrifuge shaper enables precise control of the molecular angular momentum, which in this work was varied between $N=25$ and $N=89$.

The induced magnetization of oxygen gas was detected with a small pick-up coil, centered at the location of the centrifuged volume, as shown in Fig.\ref{fig_setup}(\textbf{a}). The coil was connected to a 3 GHz bandwidth oscilloscope, which recorded the time dependent electromotive force (\textsc{emf}) $\Ea(t)$. The latter is proportional to the time derivative of the centrifuge-induced magnetization $\mathbf{M}$ perpendicular to the plane of the coil. Hereafter, we refer to the projections of $\mathbf{M}$ on the laser beam direction $\hat{x}$ and the perpendicular axis $\hat{y}$ as the \textit{longitudinal} ($\Ml$) and the \textit{transverse} ($\Mt$) magnetization, respectively. $\Ml$ was measured with a circular coil, co-axial with the centrifuge beam, whereas $\Mt$ was quantified using a tilted coil (see caption to Fig.\ref{fig_setup}). The respective magnetizations were retrieved from the recorded \textsc{emf} as
\begin{equation}\label{eq_moment}
M_{\parallel(\perp)}(t)=n_c \mu_{\parallel(\perp)}(t)=-c_{\parallel(\perp)}\int_{-\infty}^{t}\Ea(t^\prime)dt^\prime,
\end{equation}
with the proportionality coefficients $c_{\parallel(\perp)}$ calculated numerically for each coil. To find the magnetic moment per molecule, $\mu_{\parallel(\perp)}$, we divided $M_{\parallel(\perp)}$ by the number density of centrifuged molecules $n_c=\eta P/k_B T$, where $P$ and $T$ are the gas pressure and temperature, respectively, $k_B$ is the Boltzmann constant and $\eta$ is the fraction of molecules spun by the centrifuge. The intensity of our laser pulses ($\approx10^{12}$ W/cm$^{2}$) is sufficient to adiabatically spin the lowest rotational state only, dictating $\eta=0.04$ at room temperature.

To apply an external magnetic field $\mathbf{B}$, two permanent magnets were introduced along $\hat{z}$ axis, as shown in Fig.\ref{fig_setup}. Changing the gap between the magnets and their orientation, we were able to set the field strength at either 0.5 or 1~T, as well as flip its sign. Each waveform $\Ea(t)$ was averaged over 1,000 laser shots. Since the sign of the induced magnetization is defined by the direction of the molecular rotation, every experiment was repeated with the centrifuge rotation reversed, and the final signal $\Ea(t)$ was then calculated as half the difference of the two measurements. Furthermore, as discussed below, the transverse magnetization $\Mt$ must change its sign under the $B$-field inversion. Hence, it was determined from half the difference between the two \textsc{emf} signals at two opposite directions of $\mathbf{B}$.

\begin{figure}[b]
\begin{center}
    \includegraphics[width=1\columnwidth]{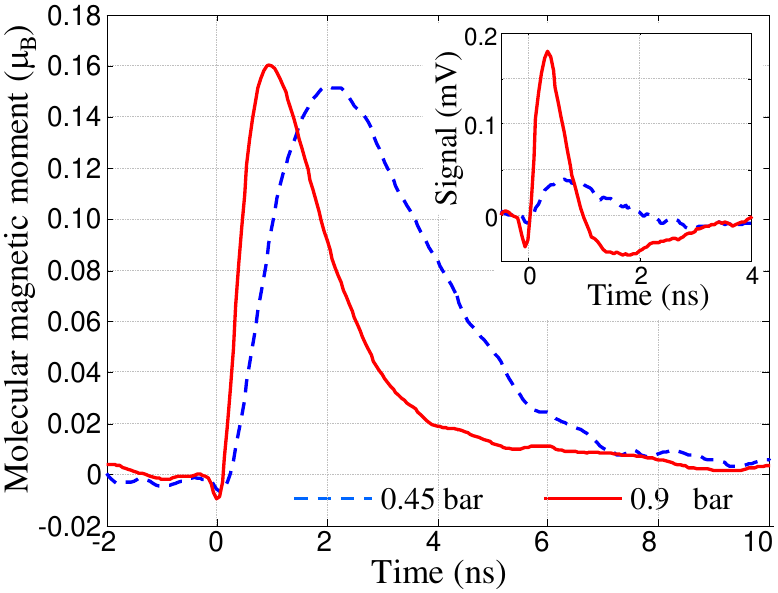}
    \caption{\label{fig_longitudinal} Centrifuge-induced longitudinal magnetic moment $\Mul$ in oxygen gas without external magnetic field, in units of Bohr magneton $\mub{}$. Inset shows raw \textsc{emf} signals $\Ea(t)$ recorded under the pressure of 0.45 bar (dashed blue) and 0.9 bar (solid red). Time zero corresponds to the arrival of the centrifuge pulse, whose duration of $\approx$30~ps was adjusted for the excitation of \otwo{} molecules to the rotational level $N=33$.}
    \end{center}
\end{figure}
\section{Longitudinal magnetization}
\subsection{Zero external field}
The observed longitudinal magnetization in the absence of the external magnetic field is shown in Fig.\ref{fig_longitudinal} for two values of the gas pressure. The detected \textsc{emf} signals are on the scale of 100~$\mu$V (see inset) and correspond to the induced magnetic moment of up to $0.16 \mub$/molecule. The clear shortening of both the rising time and the decay time of $\Mul(t)$ with increasing pressure suggests the collisional mechanism behind the induced magnetization. The spin-rotation (SR) coupling lifts the degeneracy of the three projections of oxygen's electronic spin $S$($S_N=0,\pm1$) on the angular momentum of the centrifuged molecules $\mathbf{N}$ \cite{HerzbergBook}. If the collisional spin relaxation was much faster than the rotational one, the more energetically favorable state with $S_N=+1$ would acquire higher population than that with $S_N=-1$. This would generate a non-zero gas magnetization in the direction opposite to $\mathbf{N}$, in qualitative agreement with our observations (positive sign of $\Mul$ in Fig.\ref{fig_longitudinal}).

On the other hand, the experimental absolute values of $\Ml$ are too high to be explained by this simple model. Indeed, the imbalance between the $S_N=\pm1$ states \textit{at constant $N$} is governed by the Boltzmann factor $\tanh \left(\Delta E_{N\pm1}/k_B T \right)$, with $\Delta E_{N\pm1}$ being the energy difference between the two states. For $N=33$ at room temperature, this factor of only $0.3\%$ would result in the magnetization of more than an order of magnitude smaller than the observed one. Because of the excellent agreement between the calculated and the detected transverse moment $\Mut$ (see below), we conclude that the experimentally determined magnitude of $\Mul$ is correct. We therefore associate its large value with the fact that the relaxation time scales for the electronic spin and the molecular angular momentum are not separable \cite{Milner2014b, Milner2015c}. Hence, the relaxation dynamics of the two spin components $S_N=\pm1$, occurring \textit{simultaneously with the decay of $N$}, may be quite different. A simple numerical estimate shows that a 50\% difference in the respective decay constants of the two spin components would bring the transient longitudinal magnetization close to our experimental findings. However, the validity of this model is yet to be explored.

\begin{figure}[bt]
    \begin{center}
        \includegraphics[width=1\columnwidth]{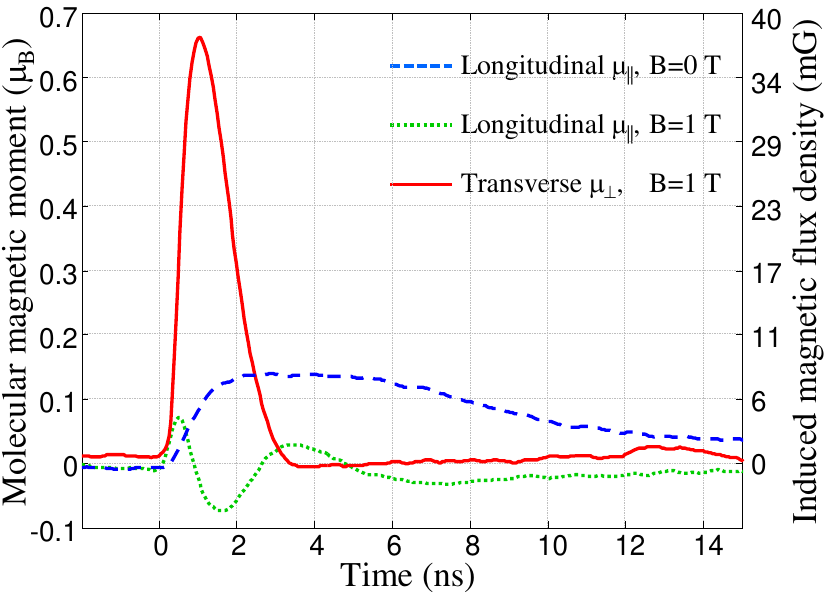}
        \caption{\label{fig_long_trans} Longitudinal (dotted green) and transverse (solid red) magnetic moments of \otwo{} molecules under the pressure of 0.5 bar, centrifuged to the rotational quantum number $N=89$, in the external magnetic field of 1~T. Longitudinal moment in the absence of $B$-field is plotted for comparison (blue dashed). Note much higher degree of rotational excitation than in Fig.\ref{fig_longitudinal} ($N=33$), which explains the longer decay time. Right vertical axis shows the corresponding values of the induced magnetic flux density.}
    \end{center}
\end{figure}
\subsection{Non-zero external field}
To study the induced magnetization further, we applied an external magnetic field along the vertical ($\hat{z}$) axis. In a recent study of the dynamics of paramagnetic oxygen super-rotors in an external magnetic field \cite{Milner2015a, Floss2015, Korobenko2015a}, it was found that the spin-rotation coupling serves as an efficient mediator between the applied $B$-field and the molecular rotation, enabling magnetic control of the rotational degree of freedom. Here, we demonstrate the effect of SR coupling on the centrifuge-induced magnetization.

The green dotted line in Fig.\ref{fig_long_trans} shows the effect of the applied field on the longitudinal magnetic moment. As expected, the latter undergoes Larmor precession, resulting in the oscillatory behavior of $\Mul(t)$. In comparison to the same signal at $B=0$ (blue dashed line), the rising edge becomes steeper in the presence of the field. While in the field-free case the rise time is of the order of a few nanoseconds and depends on the gas pressure, it becomes shorter than 0.5 ns and no longer changes with pressure when $B\neq0$. We attribute this behavior to the SR-mediated spin-flipping Raman transitions \cite{Berard1983}, which occur during the forced molecular rotation by the centrifuge (rather than during the post-centrifuge collisional thermalization) in the presence of the external $B$-field and create an imbalance between the $S_N=+1$ and $S_N=-1$ spin projections. Understanding the details of this process will require further theoretical analysis.

\section{Transverse magnetization}
Figure~\ref{fig_long_trans} shows the appearance of a strong magnetization in the transverse direction (red solid curve). It reaches $0.65\mub{}$ per every centrifuged molecule and creates a local magnetic flux density $B_\perp=\mu _0 \Mt$ of almost 40~mG in the gas under the pressure of half atmosphere (here, $\mu _0$ is the vacuum permeability). Such large value of $\Mt$ can be understood by analyzing the rotational dynamics of paramagnetic super-rotors in an external magnetic field \cite{Milner2015a, Floss2015}. An optical centrifuge drives the molecules to a state of high angular momentum $\mathbf{N}$, while an applied magnetic field generates a torque on the electronic spin $\mathbf{S}$. If the field is not strong enough to decouple the two vectors (in the case of oxygen, of order of 1~T or below), SR interaction causes the angular momentum to precess together with $\mathbf{S}$. The precession frequency depends on the mutual orientation of $\mathbf{N}$ and $\mathbf{S}$ in the following way \cite{Floss2015}:
\begin{equation}\label{eq_freq}
   \boldsymbol{\Omega}_{N,S_N}(B)=-\frac{\mub{}gS_N}{\hbar N}\mathbf{B}=-\Omega_{N}(B)S_N \hat{z},
\end{equation}
where $g$ is the electron g-factor and $\hbar$ is the reduced Planck constant. It is evident from the above expression, that the two spin components $S_N=\pm1$ precess in the opposite directions. The process is illustrated in Fig.\ref{fig_setup} by the three disk-shaped distributions of molecular axes in a high-$N$ super-rotor state, corresponding to the three possible projections of the electronic spin (short blue arrows) on the molecular angular momentum (long red arrows). After a quarter precession period (or 0.8~ns for $B=1$~T and $N=89$), the total magnetic moment in the direction perpendicular to both the initial angular momentum and the applied magnetic field reaches its maximum value of $2/3\mub{}|g|\approx 1.3\mub{}$ per molecule, close to the observed moment in Fig.\ref{fig_long_trans}. The smaller magnitude of $\Mut$ is in fact well anticipated due to the comparable time scales of the magnetic precession and rotational relaxation \cite{Milner2014b}.

\begin{figure}[bt]
    \begin{center}
    \includegraphics[width=.8\columnwidth]{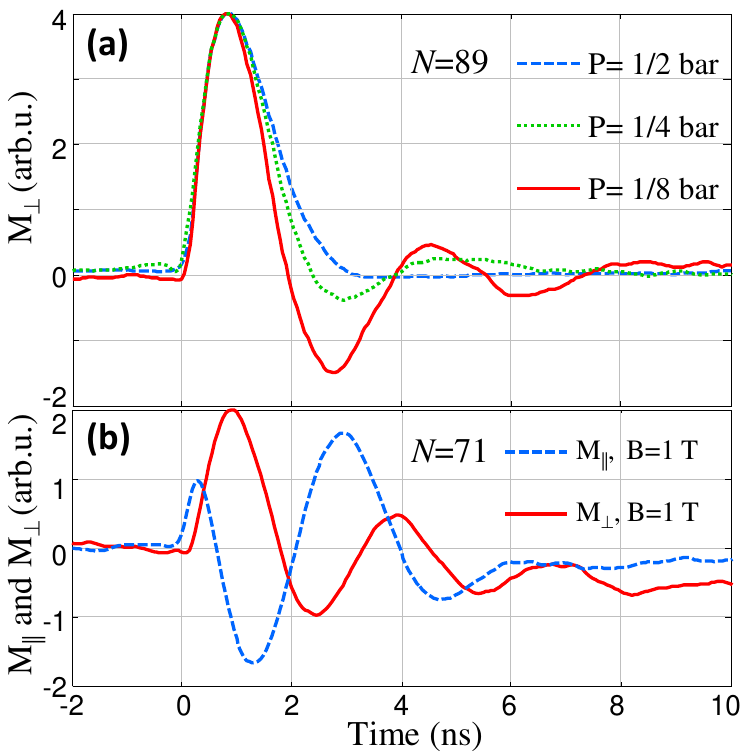}
    \caption{\label{fig_trans_p} (\textbf{a}) Dependence of the centrifuge-induced transverse magnetization on the pressure of oxygen gas in the external magnetic field of 1~T. Molecules have been excited to $N=89$. To show the precession oscillations more clearly, all signals have been normalized to the same peak value at 0.8~ns. (\textbf{b}) Longitudinal (dashed blue) and transverse (solid red) magnetization of the gas of \otwo{} molecules centrifuged to the rotational quantum number of $N=71$ in the external magnetic field of 1~T.}
    \end{center}
\end{figure}
As expected, lowering the pressure decreased the magnitude of $\Mut$ proportionally and also prolonged its lifetime, allowing to observe a larger number of oscillations, as shown in Fig.\ref{fig_trans_p}(\textbf{a}). In contrast to the longitudinal signal, no change was detected in the growth rate of $\Mut$ with pressure, which confirms its non-collisional nature. Identical precession frequencies and a $\pi /2$ phase shift between the oscillation of the longitudinal and transverse magnetic moments, clearly visible in Fig.\ref{fig_trans_p}(\textbf{b}), further corroborates their suggested origins.

\begin{figure}[bt]
    \begin{center}
    \includegraphics[width=.85\columnwidth]{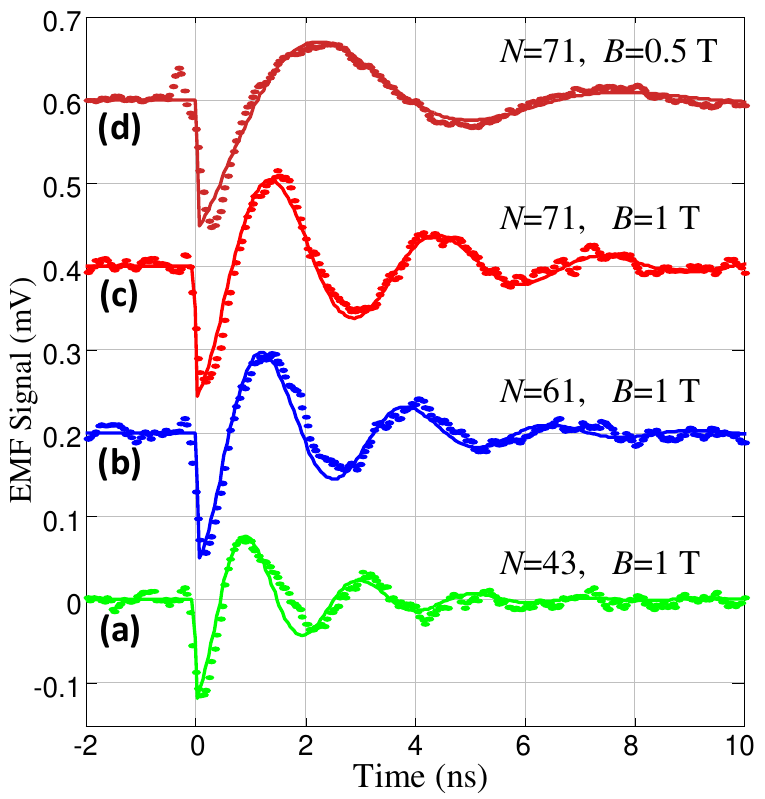}
    \caption{\label{fig_trans_nb} (\textbf{a-c}) Experimentally detected (dots) and numerically calculated (solid lines) transverse \textsc{emf} signal $\Ea(t)$ as a function of the rotational quantum number $N$ in the external magnetic field of 1~T. For clarity, each consecutive plot has been shifted up by 0.2 mV. (\textbf{d}) Same as in (\textbf{c}), but with the magnetic field strength lowered to 0.5~T.}
    \end{center}
\end{figure}
An apparent difference between the precession frequencies of the transverse magnetization in the upper and lower panels of Fig.\ref{fig_trans_p} is due to the inverse proportionality of $\Omega _{N}(B)$ on the rotational quantum number $N$, as described by Eq.\ref{eq_freq}. Fig.\ref{fig_trans_nb}(\textbf{a-c}) presents a quantitative analysis of this dependence. Different values of the molecular angular momentum were obtained by controlling the terminal frequency of the centrifuge pulse by means of adjusting its duration (for details, see \cite{Korobenko2014a}). Our experimental results (green, blue and red dots for $N=43,61$ and 71, respectively) are well described by the theoretically expected functional form (solid lines):
\begin{equation}
    \mathcal{E}(t) = -\frac{d}{dt}A_N\left(\sin\Omega_{N, +1}t+\sin\Omega_{N, -1}t\right)e^{-t/\tau_N},
\end{equation}
with the fitting parameters $A_N$ and $\tau_N$ and the two terms in the brackets corresponding to the two counter-rotating spin components $S_{N}=\pm1$. Note that using $\Omega_{N,+1}=\Omega_{N,-1}\equiv \Omega_{N}= \mub{}gB/(\hbar N)$ in accord with Eq.\ref{eq_freq} did not produce satisfactory fits, owing to the partial decoupling of $\mathbf{S}$ from $\mathbf{N}$ at high values of magnetic field \cite{Floss2015}. To account for this effect, we numerically diagonalized the spin-rotational Hamiltonian in the presence of an external magnetic field:
\begin{equation}
    \label{eq_ham}
    \hat{H}_\mathrm{SR}=\gamma\mathbf{NS}-\lambda\frac{(\mathbf{NS})^2}{N(N+1)}-g\mub{}\mathbf{SB},
\end{equation}
where $\gamma$ and $\lambda$ are the spin-rotational and the spin-spin interaction energies of oxygen, respectively \cite{BrownCarrington}. The exact frequencies $\Omega_{N,\pm1}$ were then calculated from the spectrum of the above Hamiltonian for given values of $N$ and $B$. The magnetization decay times $\tau_N$, retrieved from the fit as $3.1\pm0.6$~ns for $N=71$, $2.4\pm0.4$~ns for $N=61$ and $1.8\pm0.4$~ns for $N=43$, are in agreement with the previously found time constants for the collision-induced rotational decay \cite{Milner2014b}.
Lowering the strength of the applied magnetic field to half its value, i.e. to $B=0.5$~T, resulted in the anticipated decrease of the precession frequency, as can be seen by comparing plots (\textbf{c}) and (\textbf{d}) in Fig.\ref{fig_trans_nb}.

To summarize, we have demonstrated experimentally the ability of an optical centrifuge to induce macroscopic magnetization in a gas of paramagnetic molecules under ambient conditions. Two mechanisms have been identified and studied. The magnetic moment in the direction of the laser beam was induced by the centrifuge field alone and seems to be the result of spin-flipping collisions. The surprisingly high magnitude of this longitudinal component requires further investigation.

The second type of magnetization was observed in the transverse direction perpendicular to the centrifuge beam. It requires an external magnetic field and is well understood in the framework of the spin-rotational dynamics of paramagnetic super-rotors. Following the transfer of angular momentum from the centrifuge field to the molecular rotation, the evolution of the spin-rotational wave packet results in a transient polarization of the electronic spin. The latter process can be viewed as a rotational analogue of the spin-orbital evolution, also known to result in a transient electronic polarization in electronically excited atoms \cite{Bouchene2001, Nakajima2004}.

Extending the use of the centrifuge-induced polarization of gases to their nuclear spins is an appealing prospect, especially in the context of NMR-based medical imaging. Polarization transfer from the molecular rotation to the nuclear spin has been discussed theoretically \cite{Rakitzis2005, Sofikitis2015} and demonstrated experimentally \cite{Sofikitis2007} with low rotational excitation of HCl molecules by resonant laser pulses. Much higher degree of controllable rotational excitation with a non-resonant optical centrifuge may offer a powerful extension of this technique to a broad class of molecular species.

We would like to thank Johannes Flo{\ss} for many helpful discussions, and the electronics shop of the Department of Chemistry at UBC for invaluable continuous support.


\end{document}